# Why wave-based classical communicators can never be totally secure?

LASZLO B. KISH[+]

[+]Department of Electrical Engineering, Texas A&M University, College Station, TX 77843-3128, USA

(Version: January 12, 2007)

Recently, we proposed a classical communicator which was inspired by the Kirchhoff-loop-Johnson-like-Noise (KLJN) communicator and was claimed totally secure. Here we withdraw this claim and prove that, similarly to earlier intuitive suspicions, a wave-based classical communicator can never be totally secure.

Because the propagating wave divides the two communicators, they cannot be considered as a single organic physical system anymore. Moreover, because the signal is classical physical quantity, it is possible to measure it. The eavesdropper (Eve) can use directional couplers to measure the signal parameters of the waves propagating in both directions in the channel. Therefore, Eve can compare the wave entering into Alice's communicator and the wave leaving Alice's communicator. This is enough information to identify Alice's logic state if enough time is available. The same is true for Bob. If there is not enough time, the logic state of Alice and Bob can be estimated with certain probability and that means that there is a *non-zero information leak* toward Eve, therefore the communicator is not totally secure.

In an unpublished work by authors from the Airforce, some more fundamental problems are claimed which may indicate that these proposed solutions [1,2] are not secure at all. In the work, the authors say that, if such method could be secure than these methods could work simply via a regular emailing process by sending the wave parameters as numbers. Alice would receive the numbers from Bob and she would use the theoretical optical system response to calculate her response (outgoing signal) numbers. Bob would do the same with the numbers received from Alice. If such an email communicator would work, even with a limited-but-designable security, no physical/hardware secure communicator would ever been needed any more. We note here that this unpublished Airforce paper was written up against the KLJN communicator [3], however the paper was turned down because in the KLJN communicator the wave limit is excluded [3], and Alice and Bob are part of an organic single physical system, thus the Airforce objection does not work there.

Solution [2] has another problem. A careful check indicates that there is no propagating information. This property was hidden for the author at the beginning. Therefore we have two reasons to withdraw this paper/claim.